\documentclass[sigplan,10pt]{acmart}
\renewcommand\footnotetextcopyrightpermission[1]{}

\usepackage[english]{babel}
\usepackage{blindtext}

\usepackage{xcolor}

\usepackage{algorithm}      
\usepackage{algpseudocode}  

\renewcommand\footnotetextcopyrightpermission[1]{} 
\setcopyright{none}

\acmDOI{}

\acmISBN{}

\acmPrice{}

\begin{document}
\title[MEV Attribution at Scale]{The Origins of MEV: Systematic Attribution of Arbitrage Opportunity Creation at Scale}




\author{Andrei Seoev}
\affiliation{%
  \institution{MEV-X}
  \city{Moscow}
  \state{Russia}
}

\author{Dmitry Belousov}
\affiliation{%
  \institution{Moscow Institute of Physics and Technology}
  \city{Dolgoprudny}
  \state{Russia}
}

\author{Anastasiia Smirnova}
\affiliation{%
  \institution{Moscow Institute of Physics and Technology}
  \city{Dolgoprudny}
  \state{Russia}
}

\author{Ksenia Kurinova}
\affiliation{%
  \institution{Moscow Institute of Physics and Technology}
  \city{Dolgoprudny}
  \state{Russia}
}

\author{Aleksei Smirnov}
\affiliation{%
  \institution{MEV-X}
  \city{Moscow}
  \state{Russia}
}

\author{Denis Fedyanin}
\affiliation{%
  \institution{HSE University}
  \city{Moscow}
  \state{Russia}
}

\author{Yury Yanovich}
\affiliation{%
  \institution{Skolkovo Institute of Science and Technology}
  \city{Skolkovo}
  \state{Russia}
}


\renewcommand{\shortauthors}{X.et al.}

\begin{abstract}
    Maximal Extractable Value (MEV) represents billions of dollars in extracted value that fundamentally shapes blockchain network dynamics and participant incentives. While research has focused on MEV extraction and mitigation, we lack systematic methods to attribute MEV opportunities to their on-chain origins. This paper formalizes the MEV opportunity attribution problem and introduces a systems framework for identifying which transactions create arbitrage opportunities and quantifying their contributions. We design and evaluate four attribution methods for atomic arbitrage on EVM-compatible networks: bot-data-driven, simulation-based, coefficient-based, and Shapley-based approaches. Through large-scale retrospective analysis spanning over one million blocks on Polygon, we demonstrate that the majority of atomic arbitrage opportunities can be traced to single source transactions, validating our central hypothesis about competitive MEV markets. We quantify a highly concentrated distribution of MEV creation, where a small subset of protocols generates most opportunities, and provide comparative analysis of method trade-offs in accuracy, cost, and scalability. Our findings offer insights for protocol designers reducing MEV leakage, validators optimizing transaction ordering, and analysts measuring ecosystem health through opportunity creation.
\end{abstract}

\settopmatter{printfolios=true}
\maketitle
\pagestyle{plain}

\section{Introduction}
\label{sec:introduction}

Blockchain networks have evolved from simple payment systems to complex decentralized financial ecosystems, creating new forms of value extraction through transaction ordering. Maximal Extractable Value (MEV)--the profit that can be extracted by reordering, including, or censoring transactions within blocks--now represents billions of dollars annually across major networks, fundamentally influencing gas markets, network congestion, consensus stability, and participant incentives \cite{Daian2020,Gramlich2024}. This phenomenon exists at the intersection of network protocols, financial markets, and game theory, creating complex feedback loops that shape blockchain ecosystems. As networks scale and DeFi complexity increases, understanding and managing MEV has become a critical challenge for blockchain researchers and practitioners.

Current MEV research has predominantly focused on extraction mechanisms--how searchers identify and capture value through sophisticated strategies like arbitrage, liquidations, and sandwich attacks \cite{Torres2021FrontrunnerJA,Weintraub2022}. Measurement studies quantify extracted MEV volumes \cite{Qin2021}, while mitigation proposals range from fair ordering protocols to application-layer solutions \cite{Kelkar2020,Yang2024}. However, this extraction-centric perspective overlooks a fundamental question: \emph{which transactions create MEV opportunities in the first place?} We lack formal frameworks and methodologies for attributing MEV opportunity creation to specific on-chain actions, creating a critical gap in our understanding of the MEV supply chain. Without attribution, we cannot answer basic questions about MEV origins: Which protocols generate the most arbitrage opportunities? Which users unintentionally create value for extractors?

This paper addresses the MEV opportunity attribution problem: given an extracted MEV event (e.g., an atomic arbitrage transaction), identify which prior transaction(s) created the opportunity and quantify their contribution to the extracted value. We focus on atomic arbitrage--the most common and measurable MEV category--and investigate whether we can reliably attribute opportunities to specific source transactions using on-chain data, what methods provide the best trade-offs between accuracy and computational cost, and how MEV opportunity creation is distributed across transaction senders and protocols in practice.

A key hypothesis underpinning our work is that in competitive MEV markets, arbitrage opportunities are predominantly created by single transactions rather than complex multi-transaction sequences. This stems from the observation that searchers, operating in a highly competitive environment, extract value immediately when opportunities arise rather than waiting for optimal conditions that might require multiple coordinating transactions. This "single-source" assumption enables tractable attribution for the prevalent case while acknowledging that more complex MEV forms may require multi-transaction analysis. Our evaluation validates this hypothesis, showing that 96.7\% of atomic arbitrage opportunities can be attributed to a single source transaction.

Our work makes four key contributions:
\begin{itemize}
    \item \textbf{Formalization of MEV Attribution:} We formalize the MEV opportunity attribution problem within a blockchain state machine model, specifying inputs (blockchain state, transaction traces, MEV extraction events) and desired outputs (source transaction identification, created MEV value quantification, aggregation by sender/protocol).
    \item \textbf{Four Attribution Methods:} We design and implement four complementary attribution methods: a bot-data-driven approach that uses searcher bidding behavior as a proxy for ground truth; a simulation-based method that replays transactions to isolate state perturbations; an amplification-based technique that applies analytical models to estimate price impact; and a Shapley-based approach using cooperative game theory for fair attribution.
    \item \textbf{System Implementation and Evaluation:} We build and evaluate an attribution system for Polygon--an EVM-based blockchain--conducting large-scale retrospective analysis on 1,050,000 blocks spanning March 2026. Our evaluation demonstrates the feasibility of attribution, reveals that MEV creation is highly concentrated with a small subset of protocols responsible for most opportunities, and provides comparative performance metrics for each method.
    \item \textbf{Empirical Analysis:} We present the first systematic analysis of MEV creation distribution, identifying which protocols and users generate the most arbitrage opportunities and quantifying their contributions.
\end{itemize}

\section{Background and Scope}
\label{sec:background}

\subsection{Blockchain as a Deterministic State Machine}
\label{subsec:blockchain-model}

At a fundamental level, blockchains operate as replicated deterministic state machines. While architectural implementations vary--ranging from the UTXO model in Bitcoin~\cite{Nakamoto2008} to the account-based model in Ethereum and EVM-compatible chains like Polygon~\cite{Wood2014}--the core principle remains consistent: transactions are grouped into blocks and applied sequentially according to a consensus-determined order. In these sequential execution models, the execution of a fixed, ordered transaction sequence is strictly deterministic--the same transaction applied to the same state always yields an identical result, with no ambiguity in the transition function. Determinism holds across all nodes in the network and is enforced at the protocol level, ensuring that the observed ordering of transactions within a block serves as a fixed input to any attribution framework.

It is important to acknowledge emerging architectures that deviate from strict sequential execution. Networks such as Solana~\cite{Yakovenko2018} and TON~\cite{TONWhitepaper} employ parallel transaction execution pipelines, while sharding techniques~\cite{Wang2020ShardingSurvey} partition state across multiple chains to increase throughput. In these models, consensus still fixes the relative ordering of dependent transactions, though independent transactions may be processed concurrently. For the purposes of causal attribution, however, the critical requirement is reproducibility: any alternative execution scenario constructed by modifying the set or ordering of transactions must yield a precisely defined and verifiable state. Whether the underlying engine executes sequentially or in parallel, the consensus output provides the canonical timeline against which counterfactuals are measured.

This property is foundational to our attribution methodology. Because state transitions are fully reproducible in the sequential EVM model studied here, the blockchain provides a reliable substrate for counterfactual reasoning. The causal contribution of individual transactions to subsequent outcomes, including the emergence of MEV opportunities, can therefore be isolated and measured by replaying transaction sequences under modified conditions. We formalize this model in Section~\ref{subsec:system-model}, noting that while our empirical evaluation focuses on Polygon due to data availability and tooling support (as discussed in Section~\ref{subsec:limitations}), the formal framework applies broadly to any deterministic blockchain state machine where historical state replay is feasible.
 
\subsection{Atomic Arbitrage: Definition and Rationale}
\label{subsec:atomic-arbitrage}

Atomic arbitrage is a form of MEV extraction in which a single transaction exploits a price discrepancy across two or more liquidity pools by executing a sequence of swaps that yields a non-negative net change for each involved asset and a strictly positive profit after fees and priority bids. The transaction is atomic in the strict sense: either all constituent swaps execute successfully or none do, eliminating execution risk for the extractor. We adopt the formal identification criteria established by Vostrikov et al.~\cite{Vostrikov2025}, which classify a transaction as atomic arbitrage if and only if it satisfies three conditions: (1)~\textbf{Multi-Swap:} the transaction must contain at least two swaps ($N \geq 2$), ensuring interaction with at least two liquidity pools; (2)~\textbf{Sufficiency:} the net balance change for each asset $A \in \mathcal{A}$ must be non-negative ($\Delta(A) = \sum_{n=1}^{N} \delta_{n,A} \geq 0$, where $\delta_{n,A}$ is the change in asset $A$ at step $n$); and (3)~\textbf{Profitability:} the total profit after accounting for transaction fees $\tau$ and prioritization bids $\beta$ must be strictly positive ($\text{Profit} = \sum_{A \in \mathcal{A}} \Delta(A) \cdot P(A) - \tau - \beta > 0$, where $P(A)$ is the price of asset $A$ in the base currency, e.g., MATIC). Formally, $\text{AA} \iff (N \geq 2) \land (\forall A \in \mathcal{A}, \Delta(A) \geq 0) \land (\text{Profit} > 0)$.

\begin{figure}[t]
    \centering
    \includegraphics[width=0.85\columnwidth]{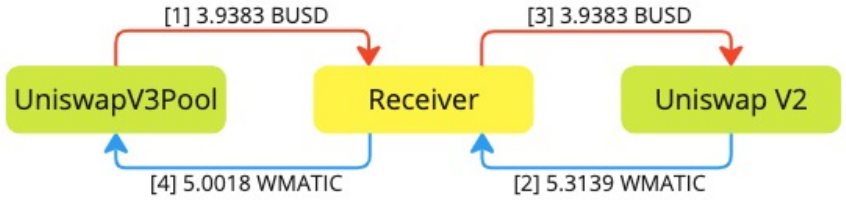}
    \caption{Example atomic arbitrage on Polygon (block 58,329,504). The arbitrageur exploits a price imbalance between Uniswap V3 (WMATIC/USDC) and Uniswap V2 (USDC/WMATIC) pools. After executing two swaps, the transaction yields a net profit of 0.3121~WMATIC (\$0.24 at execution time), satisfying all three atomic arbitrage criteria: $N=2$ swaps, $\Delta(A) \geq 0$ for all assets, and $\text{Profit} > 0$ after fees. Transaction hash: \texttt{0xc6591b9..}.}
    \label{fig:plain_arbitrage}
\end{figure}

This definition implies three key properties essential for our attribution framework. First, \textit{atomic execution guarantee}: all operations are executed within a single block; if any step fails, the entire transaction is reverted. Second, \textit{no partial execution risk}: in contrast to cross-chain MEV, where operations may complete asynchronously, atomic MEV is all-or-nothing. Third, \textit{efficiency}: these strategies exploit pricing inefficiencies between DEXs within the same blockchain. This formal definition enables precise identification of atomic MEV transactions in real-world data, which is essential for analyzing its impact on the Polygon network.

Price discrepancies of this kind arise naturally in networks where liquidity is fragmented across a large number of AMM pools that price assets independently based on their reserve balances. A single swap transaction that moves reserves in one pool does not automatically propagate to others, creating transient imbalances that persist until arbitraged away. On Polygon, where a new block is produced approximately every two seconds and each block may contain hundreds of transactions, such imbalances arise frequently and are typically resolved within a single block.

Atomic arbitrage is the most tractable starting point for an MEV attribution framework for several reasons. Among all MEV categories, it occurs with the highest frequency on-chain and produces outcomes that are directly and fully observable from transaction data. The value captured in each extraction event can be computed deterministically from pool states alone, requiring no off-chain inputs or external price references. From a causal standpoint, the structure is well-contained: the opportunity emerges from state perturbations introduced by preceding transactions and is consumed by the arbitrage transaction itself, making the causal chain both bounded and analyzable. The absence of any off-chain coordination further ensures that the entire sequence of events is captured within the on-chain record, making attribution tractable within the state machine model introduced in Section~\ref{subsec:blockchain-model}.
 
\subsection{Core Definitions}
\label{subsec:definitions}
 
The following definitions establish the conceptual vocabulary used throughout this paper. They
form a causal chain central to our analysis: individual transactions alter blockchain state,
state changes give rise to exploitable opportunities, those opportunities are captured by
arbitrageurs, and attribution seeks to trace that chain back to its origin.
 
\smallskip
\noindent\textbf{MEV Extraction.} MEV extraction refers to the act of capturing profit by
strategically reordering, including, or censoring transactions within a block. In this work,
extraction is represented by an executed atomic arbitrage transaction that realizes a positive
profit by exploiting an existing price imbalance across liquidity pools.
 
\smallskip
\noindent\textbf{MEV Opportunity.} An MEV opportunity is a blockchain state in which a
profitable arbitrage route exists and can be executed by a suitably constructed transaction.
Opportunities emerge when transactions introduce state perturbations that shift pool prices
away from equilibrium, and persist until an arbitrageur acts on them.
 
\smallskip
\noindent\textbf{State Delta.} A state delta is the change in blockchain state produced by
the execution of a single transaction; we refer to this equivalently as a state
perturbation throughout the causal analysis of Section~\ref{sec:problem}. In the context 
of this work on Polygon, the relevant component of a state delta is the shift in liquidity 
pool reserves--and consequently pool prices--induced by a swap transaction on EVM-compatible 
AMMs (e.g., Uniswap V2/V3, QuickSwap). State deltas are the mechanism through which individual 
transactions create MEV opportunities; a transaction \emph{amplifies} an existing opportunity 
if it increases the profit of an already-profitable arbitrage route, a distinction formalized 
in the attribution methods of Section~\ref{sec:methodology}. We operationalize state deltas 
via deterministic replay on an archive node, enabling precise counterfactual simulation of 
pool states at arbitrary positions in the transaction sequence.

\section{Problem Formulation}
\label{sec:problem}

\subsection{Goal: Attribute MEV Opportunity to a Source Transaction}
\label{subsec:goal}

The fundamental objective of this work is to resolve the provenance of Maximal Extractable Value (MEV) by establishing a causal link between value extraction and value creation. Specifically, given an observed atomic arbitrage transaction $T_{arb}$ that extracts profit $\Pi$, we seek to identify the preceding transaction or set of transactions $T_{src}$ that created the price disbalance enabling that profit. This attribution problem is distinct from MEV extraction; rather than determining how to capture value, we aim to determine which on-chain actions introduced the opportunity and quantify their contribution to the total extracted value. In competitive markets such as Polygon, we hypothesize that the majority of atomic arbitrage opportunities are triggered by a single source transaction rather than complex multi-transaction sequences. This single-source assumption simplifies the attribution task for the prevalent case, allowing us to map extracted profit back to specific on-chain actions with high confidence. Our goal is to produce a mapping $\mathcal{A}$ where each MEV extraction event is linked to its causal source, quantifying the contribution $\phi_i$ of each source transaction $T_i$ to the total extracted value $\Pi$.

\subsection{System Model and Inputs}
\label{subsec:system-model}

Building on Section~\ref{subsec:blockchain-model}, we model a block $B$ as an ordered transaction sequence $(T_1, \dots, T_n)$ with sequential state evolution $S_k = \Sigma(S_{k-1}, T_k)$. Since state persists across blocks, a source transaction $T_i$ may precede $T_{arb}$ within the same block or up to $D=100$ earlier blocks.

Our system requires three inputs: (1)~\textbf{Blockchain State}: pool reserves and prices at $S_0$ (via archive node); (2)~\textbf{Transaction Trace}: ordered transactions with calldata/logs for deterministic replay; (3)~\textbf{MEV Event}: the arbitrage transaction $T_{arb}$ at index $k$, its execution path, and extracted profit~$\Pi$.

\subsection{Desired Outputs and Scope}
\label{subsec:desired-outputs}

Let $\mathcal{M}(S, T_{arb})$ denote the profit from executing $T_{arb}$ in state $S$. Total extracted profit is $\Pi = \mathcal{M}(S_{k-1}, T_{arb})$, where $S_{k-1}$ is the state after all preceding transactions. Candidate sources are $C = \{T_i \mid i < k \land T_i \text{ interacts with pools in } T_{arb}\}$. We seek a value distribution $\{\phi_i\}_{T_i \in C}$ satisfying $\sum_{T_i \in C} \phi_i + \phi_{base} = \Pi$, where $\phi_{base} = \mathcal{M}(S_0, T_{arb})$.

For each event, we output: (1)~\textbf{Source ID}: hash of the primary contributing transaction; (2)~\textbf{Created MEV}: attributed value $\phi_i$; (3)~\textbf{Aggregation meta} sender/protocol tags for high-level analysis.

We limit scope to \textbf{Atomic Arbitrage}~\cite{Vostrikov2025}: transactions with $\geq 2$ swaps, non-negative net balance changes for all assets, and positive profit after fees/bids. Other MEV categories (liquidations, sandwich attacks) require distinct causal models and are excluded.

\section{Attribution Methodology}
\label{sec:methodology}

\subsection{Overview and Pipeline}
\label{subsec:pipeline}

Our attribution pipeline processes blockchain blocks in sequential order, beginning with the identification of atomic arbitrage transactions according to the criteria established in Section~\ref{subsec:desired-outputs}. Upon flagging an arbitrage transaction $T_{arb}$, the system isolates the set of preceding transactions within the same block (and up to $D$ preceding blocks) that interacted with the liquidity pools involved in the arbitrage route. This candidate set, denoted $C$, is then subjected to one or more attribution methods to determine causal responsibility and quantify value contribution.

We conceptualize blockchain state evolution as a sequence of \emph{state perturbations} induced by transaction execution. Each transaction modifies the deterministic state of the blockchain, including liquidity pool reserves, token balances, and smart contract storage. An MEV opportunity arises when a sequence of such perturbations creates a price disbalance that can be exploited through atomic arbitrage. Our attribution framework treats the problem as one of causal inference: given an observed profit extraction event, we seek to identify which prior state perturbation was primarily responsible for enabling that extraction. This perspective aligns with counterfactual reasoning in systems analysis, where we ask what would have happened to the arbitrage profit had a particular transaction not been executed or had it been executed at a different position in the block.

The pipeline architecture is intentionally modular, permitting different attribution methods to execute in parallel for comparative analysis. Post-processing steps aggregate results across the entire dataset to compute distributional statistics and identify patterns in MEV opportunity creation. The entire process operates offline in a retrospective manner, prioritizing analytical accuracy over real-time latency, which enables the use of computationally intensive simulation techniques that would be impractical in a live trading environment. We formalize the pipeline as a function $\mathcal{A}(B, T_{arb})$ mapping a block $B$ and an arbitrage transaction $T_{arb}$ to a set of pairs associating source transactions with their attributed value contributions.

\subsection{Method Summary and Trade-offs}
\label{subsec:method-summary}

We implement four distinct methodological approaches to address the MEV attribution problem, each offering different trade-offs between analytical rigor, computational efficiency, and data requirements. These methods range from leveraging observed behavior of production MEV searchers to applying formal cooperative game theory. For brevity, we define the profit function $F(S) = \mathcal{M}(S, T_{arb})$ representing the profit achievable by executing $T_{arb}$ in blockchain state $S$.

\textbf{Real-time vs. retrospective methods.} A key distinction among our methods is their temporal orientation. Bot-data attribution operates in \emph{real-time}, capturing searcher \emph{intents} and predictions about which transactions will create opportunities. In contrast, simulation, coefficient, and Shapley methods are \emph{retrospective}, analyzing what \emph{actually happened} after blocks are finalized. This distinction is important: bot data reflects market expectations and competitive intelligence (what searchers \emph{believed} was causal), while the other methods provide ground-truth causal analysis (what \emph{was} causal). In our evaluation, we use bot data as an external validation signal rather than primary attribution, as retrospective methods offer complete coverage and deterministic reproducibility.

Table~\ref{tab:method-characteristics} summarizes the key characteristics of each method. In practice, we recommend a hybrid workflow: use the coefficient-based method for rapid initial screening, apply simulation-based attribution as the primary method for most opportunities, and reserve Shapley-based computation for edge cases where the faster methods disagree or where multiple transactions appear to have substantial marginal impacts. The bot-data method, when available, serves as an external validation layer that can confirm or challenge the conclusions of the simulation-based approach.

\begin{table*}[t]
\centering
\caption{Comparative characteristics of the four attribution methods. $|C|$ = candidate set size; $m$ = transactions in edge-to-arbitrage range; $N$ = Monte Carlo samples.}
\label{tab:method-characteristics}
\small
\renewcommand{\arraystretch}{1.05}
\setlength{\tabcolsep}{0.7em}
\begin{tabular}{@{}l@{\hspace{0.8em}}p{2.2cm}@{\hspace{0.8em}}p{4.5cm}@{\hspace{0.8em}}p{2.2cm}@{\hspace{0.8em}}p{3.3cm}@{}}
\hline
\textbf{Method} & \textbf{Complexity} & \textbf{Data Requirements} & \textbf{Scalability} & \textbf{Recommended Use} \\
\hline
Bot-data & $O(1)$ after model loading & Bidding logs, route predictions & Limited by bot coverage & Validation \\
Coefficient & $O(1)$ & Pool prices from logs & $|C|\lesssim 1000$ & Initial filtering \\
Simulation & $O(\log|C| + m)$ & Historical states, execution engine & $|C|\lesssim 200$ & Primary attribution \\
Shapley (exact) & $O(2^{|C|})$ & Full simulation + enumeration & $|C| < 20$ & Ground truth \\
Shapley (MC) & $O(N \cdot |C|)$ & Full simulation + sampling & $|C| \leq 50$ & Approximate ground truth \\
\hline
\end{tabular}
\end{table*}

\subsection{Attribution Methods}
\label{subsec:methods}

\subsubsection{Bot Data-Based Attribution}
\label{subsubsec:bot-data}

This method leverages the behavior of production MEV searcher bots as a proxy for ground truth attribution. In highly competitive MEV markets, searchers employ lightweight machine learning models to identify opportunity-creating transactions and formulate profitable arbitrage routes in real time.

\textbf{Model architecture.} Our implementation uses a reinforcement learning (RL) agent with a two-component architecture: (1) a graph neural network (GNN) encoder that processes the liquidity pool topology and current reserves, producing a state embedding; and (2) a multi-layer perceptron (MLP) value head that predicts the expected profit of executing a candidate arbitrage route given that embedding. The agent is trained via proximal policy optimization (PPO) on historical Polygon data, with the reward function defined as realized arbitrage profit minus gas costs and priority bids. The model is designed for low-latency inference ($<10$ ms per candidate) to operate in real-time mempool conditions.

\textbf{Inference and attribution.} During operation, the agent monitors pending transactions and, for each candidate source transaction $T_i$, computes the optimal arbitrage route and maximum bid that would yield positive expected profit. When the bot submits a bid on a route triggered by $T_i$, we interpret this as strong evidence that $T_i$ was identified as the primary opportunity creator. Formally, let $D_{bot}$ denote the dataset of bot bidding decisions, where each entry contains a tuple $(T_{arb}, T_{target}, \text{bid\_amount}, \text{predicted\_route})$. We define the attribution indicator $I_{bot}(T_i) = 1$ if $\exists (T_{arb}, T_{target}, \text{bid}, \cdot) \in D_{bot}$ with $T_{target} = T_i$ and $\text{bid} > \theta$ (a minimal contention threshold), and $0$ otherwise.

\textbf{Practical considerations.} This method benefits from incorporating the collective intelligence of live market participants who have financial incentives to correctly identify opportunity sources. However, it is inherently limited to opportunities that were actually observed and contested by the bot infrastructure (from February 2026 onwards in our dataset), and it requires access to detailed bidding data or mempool surveillance feeds that may not be publicly available. Critically, bot data reflects \emph{intentions} rather than outcomes: it shows what searchers \emph{believed} was causal at submission time, not what retrospective analysis confirms.

\subsubsection{Simulation-Based Attribution}
\label{subsubsec:simulation}

The simulation-based method employs counterfactual execution to isolate the causal impact of specific transactions on arbitrage profitability. All profit values are normalized and expressed in MATIC on the Polygon network to ensure consistent comparison across diverse token pairs. The method proceeds in three distinct phases, formalized in Algorithm~\ref{alg:simulation}.

\textbf{Phase 1: Candidate filtering.} We eliminate from consideration all transactions that do not interact with any of the liquidity pools involved in the arbitrage opportunity, thereby reducing the candidate set to only those transactions that could plausibly have affected the relevant prices.

\textbf{Phase 2: Binary search for edge transaction.} We apply a binary search procedure backwards from $T_{arb}$ to identify the \emph{edge transaction} $T_{edge}$, defined as the earliest position in the sequence (searching backwards) at which executing the arbitrage transaction yields a profit that drops to at most 5\% of the original observed profit $\Pi$. The search begins within the current block; if no such transaction is found, it extends backwards into preceding blocks, up to a maximum depth of $D$ blocks. The parameter $D$ is blockchain-dependent: it should scale with block time, searcher competition, and pool volatility. For Polygon (2-second blocks, high competition), we set $D_{\max} = 100$ as a conservative upper bound. In our empirical evaluation across 2M blocks, we observed that 99.3\% of attributable opportunities have $T_{edge}$ within 7 blocks of $T_{arb}$, suggesting that $D=100$ provides substantial headroom for edge cases. If attribution points to the pre-block state $S_0$ (no responsible transaction found within the search window), we record the opportunity as pre-existing and not attributable to any in-block transaction.

\textbf{Phase 3: Backward impact calculation for source identification.} 
Starting from $T_{arb}$, we compute the marginal impact of each transaction in the range 
$[T_{edge}, T_{arb}]$ by traversing backwards. Let $\Pi(T_i)$ denote the accumulated MEV 
available if $T_{arb}$ were executed immediately after $T_i$. The impact of transaction 
$T_i$ is defined as the difference in accumulated MEV before and after its execution:
\[
\text{Imp}_i = \Pi(T_{i+1}) - \Pi(T_i)
\]
We traverse backwards from $T_{arb}$ to $T_{edge}$, computing $\text{Imp}_i$ for each 
transaction. The source transaction is selected as the one with maximum positive impact:
\[
T_{src} = \arg\max_{T_i \in [T_{edge}, T_{arb}]} \text{Imp}_i
\]
In cases where multiple transactions yield the same maximal impact, we select the one 
closest to $T_{arb}$ in the transaction ordering (tie-breaking rule: 
$\arg\min(\text{pos}(T_{arb}) - \text{pos}(T_i))$).

This method provides strong causal evidence through direct counterfactual simulation, but it requires access to historical blockchain states and the ability to simulate transaction execution with high fidelity. The computational cost scales logarithmically with the search depth due to the binary search optimization, followed by a linear scan in the worst case.

\begin{algorithm}[t]
\caption{Simulation-Based Attribution}
\label{alg:simulation}
\begin{algorithmic}[1]
\Require Arbitrage tx $T_{arb}$, candidates $C$, profit func $\Pi$, depth $D=100$
\Ensure Source tx $T_{src}$ or \textsc{None}
\State \textbf{Phase 1: Filter} $C \gets \{T_i \in C \mid T_i \text{ interacts with pools in } T_{arb}\}$
\State \textbf{Phase 2: Binary search for edge} $T_{edge}$
\State $low \gets \max(0, \text{pos}(T_{arb}) - D \cdot B_{\text{size}})$
\State $high \gets \text{pos}(T_{arb}) - 1$
\While{$low < high$}
    \State $mid \gets \lfloor(low + high) / 2\rfloor$
    \State $\Pi_{mid} \gets \text{simulate}(S_0, T_1, \dots, T_{mid}, T_{arb})$
    \If{$\Pi_{mid} \leq 0.05 \cdot \Pi(T_{arb})$} \Comment{Find 95\% profit drop}
        \State $high \gets mid$
    \Else
        \State $low \gets mid + 1$
    \EndIf
\EndWhile
\State $T_{edge} \gets T_{low}$
\If{$\Pi(T_{edge}) < 0.05 \cdot \Pi(T_{arb})$} \Return \textsc{None} \EndIf
\State \textbf{Phase 3: Backward impact calculation}
\State $best\_impact \gets 0$; $T_{src} \gets \textsc{None}$
\For{$T_i$ in reverse order from $T_{arb}-1$ down to $T_{edge}$}
    \State $\Pi_i \gets \text{simulate}(S_0, T_1, \dots, T_i, T_{arb})$
    \State $\text{Imp}_i \gets \Pi_{i+1} - \Pi_i$ \Comment{Marginal impact}
    \If{$\text{Imp}_i > best\_impact$}
        \State $best\_impact \gets \text{Imp}_i$; $T_{src} \gets T_i$
    \EndIf
\EndFor
\State \Return $T_{src}$
\end{algorithmic}
\end{algorithm}

\begin{figure}[t]
    \centering
    \includegraphics[width=\columnwidth]{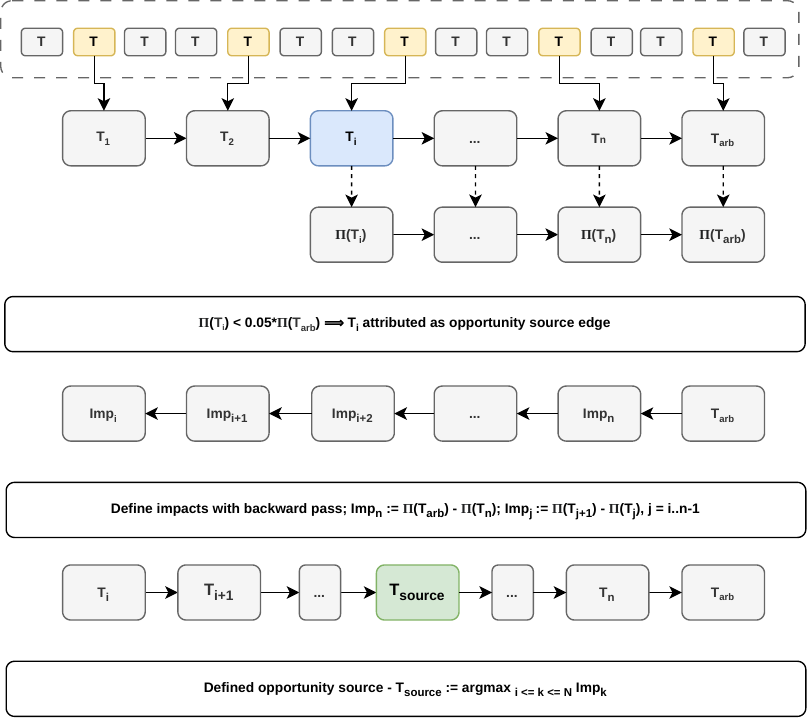}
    \caption{Simulation-based attribution pipeline. (1) Filter transactions by pool intersection 
    (yellow). (2) Binary search backwards to find edge transaction $T_{edge}$ where profit drops 
    below 5\% threshold (blue). (3) Compute marginal impacts via backward pass; select source 
    transaction with maximum impact (green).}
    \label{fig:simulation_pipeline}
\end{figure}

\subsubsection{Coefficient-Based Attribution}
\label{subsubsec:coefficient}

Also referred to as the K-value method, this approach analyzes the mathematical impact of transactions on pool pricing coefficients without requiring full state simulation. For a given arbitrage direction through a cycle of liquidity pools, we compute the price multiplier coefficient $k$, which represents the theoretical profitability of exchanging an infinitesimally small value through that cycle under the current pool reserves. If $k > 1$, the route is potentially profitable in the absence of slippage and fees.

We then examine the sequence of transactions within the block to identify which transaction caused the greatest marginal increase in the coefficient $k$. Let $k(S)$ denote the price multiplier coefficient computed from pool reserves in state $S$. We attribute the opportunity to the transaction $T_i$ that maximizes the marginal impact $\delta_{k_i} = k(S_i) - k(S_{i-1})$, where $S_i$ is the state after executing transaction $T_i$. Formally, $T_{src} = \arg\max_{T_i \in C} (k(S_i) - k(S_{i-1}))$. Algorithm~\ref{alg:coefficient} provides the complete procedure.

This method is computationally efficient because it requires only pool price data that can be extracted from transaction execution logs or callback events, rather than full historical state reconstruction. However, it does not account for liquidity depth, price slippage, or transaction fees, making it less accurate for large-volume arbitrage opportunities where non-linear price impacts are significant. The coefficient-based method is best suited for initial filtering or for scenarios where computational resources are severely constrained.

\begin{algorithm}[t]
\caption{Coefficient-Based Attribution (K-Value Method)}
\label{alg:coefficient}
\begin{algorithmic}[1]
\Require Arbitrage transaction $T_{arb}$, candidate set $C$, pool reserve function $R$
\Ensure Source transaction $T_{src}$
\State Compute arbitrage route pools $\mathcal{P} = \{P_1, \dots, P_m\}$ from $T_{arb}$
\State $k_{base} \gets \text{compute\_k}(R(S_0), \mathcal{P})$ \Comment{Base coefficient at $S_0$}
\State $best\_delta \gets 0$; $T_{src} \gets \textsc{None}$
\For{$T_i$ in chronological order in $C$}
    \State $S_i \gets \text{apply\_tx}(S_{i-1}, T_i)$ \Comment{Update state (reserves only)}
    \State $k_i \gets \text{compute\_k}(R(S_i), \mathcal{P})$
    \State $\delta_{k_i} \gets k_i - k_{i-1}$ \Comment{Marginal coefficient change}
    \If{$\delta_{k_i} > best\_delta$ \textbf{or} ($\delta_{k_i} = best\_delta$ \textbf{and} $\text{pos}(T_i) > \text{pos}(T_{src})$)}
        \State $best\_delta \gets \delta_{k_i}$; $T_{src} \gets T_i$
    \EndIf
\EndFor
\State \Return $T_{src}$
\end{algorithmic}
\end{algorithm}

\subsubsection{Shapley-Based Attribution}
\label{subsubsec:shapley}

The Shapley-based method applies cooperative game theory to fairly attribute profit when multiple transactions may have contributed to an MEV opportunity. We model the preceding candidate set $C$ as players in a cooperative game, where the value function $V(S)$ for any subset $S$ of transactions is defined as the arbitrage profit achievable after executing exactly the transactions in $S$. The Shapley value $\phi_i$ for each transaction $T_i$ is then computed by averaging its marginal contribution across all possible permutations of transaction subsets:

\[
\phi_i = \sum_{S \subseteq C \setminus \{T_i\}} \frac{|S|! (|C| - |S| - 1)!}{|C|!} \left( V(S \cup \{T_i\}) - V(S) \right).
\]

While this method provides the most theoretically rigorous attribution for multi-source opportunities, its exact computation has exponential complexity in the number of $C$. To manage this complexity, we first eliminate any transactions that do not interact with the relevant liquidity pools, thereby reducing the candidate set. For larger candidate sets ($|C| \geq 20$), we employ Monte Carlo approximation techniques: rather than enumerating all possible subsets, we sample a fixed number of random permutations of the $C$ and estimate the Shapley value as the average marginal contribution across these samples. Algorithm~\ref{alg:shapley} formalizes the Monte Carlo procedure. The total extracted profit is then decomposed as the sum of a base component available from the initial block state plus the Shapley values attributed to each candidate transaction.

This method serves as our computational ground truth for evaluating the other methods, though its infrastructure requirements and execution time make it impractical for real-time or large-scale retrospective analysis without approximation. Convergence analysis shows that 1000 samples suffice for estimates to stabilize within 5\% of asymptotic values for typical candidate set sizes in our dataset.

\begin{algorithm}[t]
\caption{Shapley-Based Attribution (Monte Carlo Approximation)}
\label{alg:shapley}
\begin{algorithmic}[1]
\Require Arbitrage transaction $T_{arb}$, candidate set $C$, profit function $V$, samples $N$
\Ensure Shapley values $\{\phi_i\}_{T_i \in C}$
\State Initialize $\phi_i \gets 0$ for all $T_i \in C$
\For{$j \gets 1$ to $N$}
    \State $\sigma \gets \text{random\_permutation}(C)$ \Comment{Random ordering of candidates}
    \State $S \gets \emptyset$
    \For{$T_i$ in order $\sigma$}
        \State $\text{marginal} \gets V(S \cup \{T_i\}) - V(S)$
        \State $\phi_i \gets \phi_i + \text{marginal}$
        \State $S \gets S \cup \{T_i\}$
    \EndFor
\EndFor
\State $\phi_i \gets \phi_i / N$ for all $T_i \in C$ \Comment{Average over samples}
\State $T_{src} \gets \arg\max_{T_i \in C} \phi_i$ \Comment{Primary source}
\State \Return $\{\phi_i\}_{T_i \in C}$, $T_{src}$
\end{algorithmic}
\end{algorithm}

\section{Evaluation}
\label{sec:evaluation}

\subsection{Experimental Setup}
\label{subsec:experimental-setup}

Our evaluation uses historical block data from the Polygon network, an EVM-compatible blockchain with high DeFi activity and frequent atomic arbitrage opportunities~\cite{Vostrikov2025}. We collected data spanning two distinct periods:

\textbf{Large-scale attribution analysis (March 2026).} Blocks 83,770,001 through 84,820,000, corresponding to March 2--29, 2026. This period includes both normal operating conditions and several episodes of elevated volatility, providing a diverse testbed for attribution methods. We identified 360,026 atomic arbitrage transactions using the three-criteria definition from prior work~\cite{Vostrikov2025} (adopted in Section~\ref{subsec:atomic-arbitrage}), with total extracted MEV volume of \$334,799 (all values reported in USD at execution-time oracle prices; internal computation in MATIC). This dataset forms the basis for our distributional analysis of MEV creation by protocol (Section~\ref{subsec:results}).

\textbf{Method comparison study (February 2026).} Blocks 82,546,747 through 82,567,395, corresponding to a near 12-hour window on February 4, 2026. This focused dataset contains 2,526 atomic arbitrage events and was used for comparative evaluation of all four attribution methods (bot-data, simulation, coefficient, Shapley), as well as validation of the single-source hypothesis. The shorter window was selected to enable exhaustive Shapley computation for ground-truth validation while maintaining sufficient diversity in transaction patterns.

We processed blocks using a modified Geth archive node configured to expose historical state snapshots and transaction traces. Our attribution pipeline was implemented in Rust for performance-critical components, with Python wrappers for statistical analysis and visualization. Experiments were conducted on a cluster of machines with 32 physical cores (Intel Xeon Platinum, 2.5--3.5 GHz) and 128 GB RAM, comparable to AWS m5.8xlarge instances. Each attribution method executed in isolated containers to ensure reproducible resource allocation.

\subsection{Ground Truth and Validation Strategy}
\label{subsec:ground-truth}

Establishing ground truth for MEV attribution is challenging as causal relationships are not directly observable on-chain. We employ triangulation on the February 2026 dataset (2,526 events): (1)~\textbf{Bot consensus}: consistent targeting by independent searchers indicates correct attribution (coverage: 38.4\%, limited to our bot's real-time mempool visibility and missing last-in-block atomic arbitrage by design); bot data reflects searcher \emph{intents} (real-time predictions about which transactions will create opportunities) rather than retrospective causal analysis, which systematically limits agreement with methods that analyze finalized blocks. (2)~\textbf{Shapley ground truth}: exact values via exhaustive enumeration for $|C|<20$ (coverage: 12.0\%); the remaining events split into (a)~2\% with $|C|\geq 20$ (computationally intractable for exact computation) and (b)~23\% attributed to $S_0$ $S_0$ (opportunities originating prior to our search window), which account for $<4\%$ of total profit. Monte Carlo Shapley (1000 samples) provides attribution for all events (100\% coverage) and agrees with exact Shapley by construction where both are feasible. For agreement computation with Shapley, attributions to the $S_0$ (i.e., "no in-block source") are counted as matches when multiple methods concur on this outcome. (3)~\textbf{Manual case studies}: expert review of 200 stratified events for plausibility assessment. A method is correct if its attribution matches consensus of $\geq 2$ sources--a conservative benchmark acknowledging inherent uncertainty.

\subsection{Shapley Attribution: Examples and Convergence}
\label{subsec:shapley-results}

\begin{figure}[t]
    \centering
    \includegraphics[width=\columnwidth]{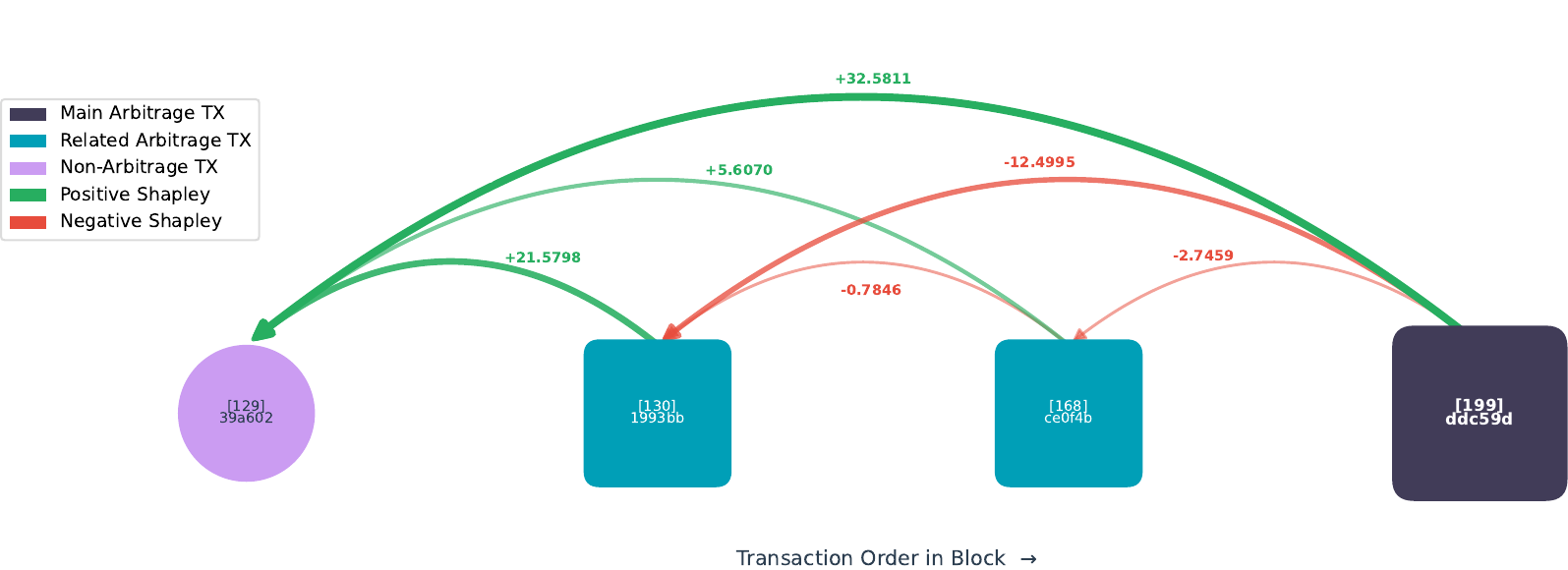}
    \caption{Shapley attribution for an arbitrage event (block 82,563,006). Positive values indicate opportunity creation; negative values indicate profit consumption by competing arbitrageurs. The non-arbitrage transaction at index 129 is the primary source (+32.58 MATIC).}
    \label{fig:shapley-example}
\end{figure}

\textbf{Shapley illustration.}
To illustrate the behavior of Shapley-based attribution, Figure~\ref{fig:shapley-example} presents a representative case from block 82,563,006 (within the February comparison window). Four candidate transactions interact with the arbitrage route: two competing arbitrageurs (indices 130, 168), one non-arbitrage swap (index 129), and the executed arbitrage transaction itself (index 199). The Shapley values reveal that the non-arbitrage transaction at index 129 contributes $+32.58$ MATIC to the opportunity--the largest positive attribution--while the competing arbitrageurs exhibit negative contributions ($-12.50$, $-2.75$ MATIC), reflecting their consumption of available profit. This example demonstrates how Shapley attribution fairly distributes value among interdependent transactions, capturing both opportunity creation and competition effects.

\textbf{Single-source vs. multi-source prevalence.} Across our February 2026 dataset (2,526 atomic arbitrage events), 83 events (3.3\%) exhibit tied maximum Shapley values, indicating genuine multi-source opportunity creation. Among these, 71 events have 2 tied sources, 8 have 3 tied sources, and 2 events each have 4 or 6 tied sources. Notably, 42 of these 83 multi-source cases (50.6\%) have equal Shapley values for all positive contributors, typically corresponding to ``blind'' last-in-block arbitrage that interacts with pools closing arbitrage cycles within the same block. The remaining 33 are last-in-block arbitrage with cascades of interdependent arbitrageurs before it. For the predominant single-source cases (96.7\%), one transaction accounts for $>70\%$ of total positive Shapley value, validating our simulation-based method's focus on identifying the primary contributor.

\textbf{Single-source hypothesis validation.} Our methodology assumes that competitive MEV markets predominantly produce single-source opportunities. Empirical results support this: 96.7\% of events have one transaction accounting for >70\% of positive Shapley value, with the next largest contributor typically <5\%. This pattern holds across protocol categories and volatility regimes, justifying simulation-based attribution's focus on primary contributors.

\begin{figure*}[t]
    \centering
    \includegraphics[width=0.8\linewidth]{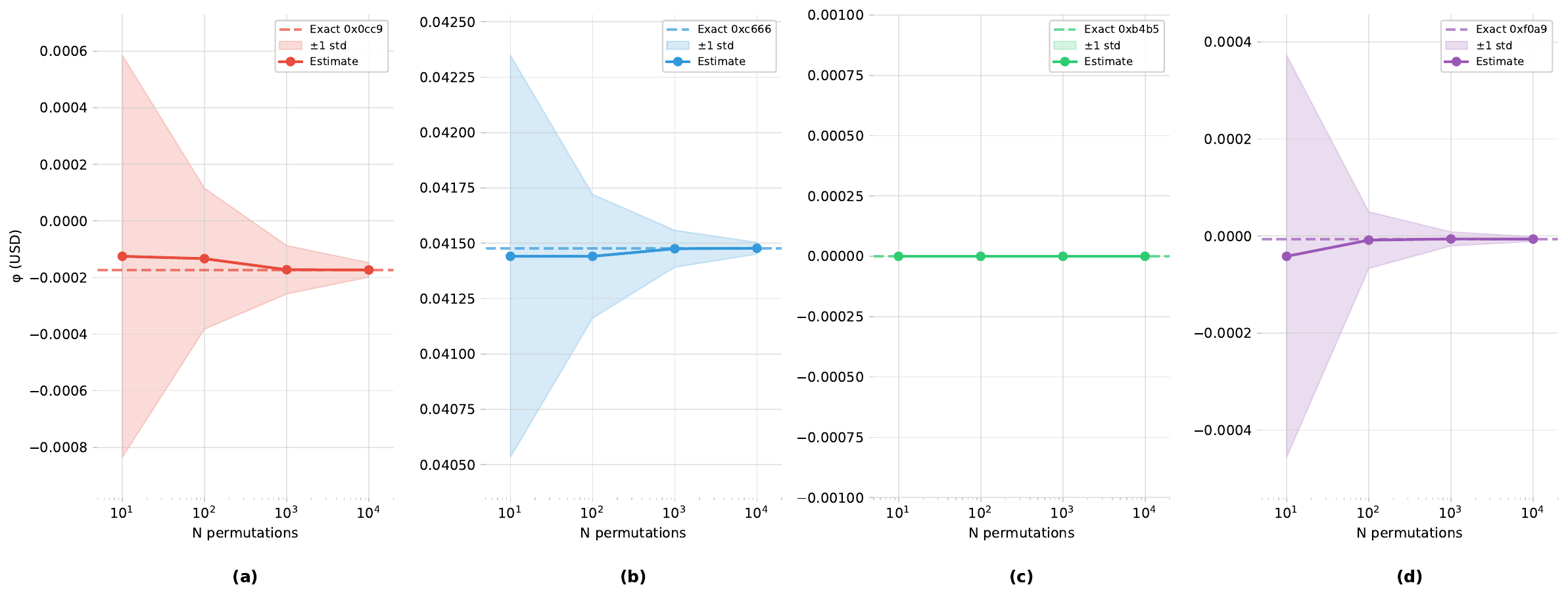}
    \caption{Monte Carlo Shapley convergence for transaction \texttt{0xb1f2a5bb..} (block 82,554,874). Four subplots show convergence for candidates with varying Shapley values (one near-zero, one exactly zero, two negative). Exact Shapley values shown as horizontal dashed lines; Monte Carlo estimates (mean $\pm$1 std over 100 runs) shown as points with shaded regions. Estimates stabilize within 5\% after $\sim$500 samples.}
    \label{fig:shapley-convergence}
\end{figure*}

\textbf{Monte Carlo convergence.} For larger candidate sets, we employ Monte Carlo approximation. Figure~\ref{fig:shapley-convergence} illustrates convergence for transaction 0xb1f2a5bb... in block 82,554,874, which has 16 candidate transactions with potentially non-zero Shapley values. Exact computation would require evaluating $2^{16} = 65,536$ subsets--computationally feasible for validation but prohibitive at scale. The figure shows four representative candidates spanning the range of Shapley values ($\phi_{0xc666} = 0.0415$, $\phi_{0x0cc9} = -0.00017$, $\phi_{0xb4b5} = 0.0$, $\phi_{0xf0a9} \approx -6.5 \times 10^{-6}$). Estimates stabilize within 5\% of their asymptotic values after approximately 500 samples, justifying our default setting of 1000 samples. This convergence behavior enables Shapley-based attribution at scale while preserving theoretical fairness guarantees.

\begin{figure*}[t]
    \centering
    \includegraphics[width=\linewidth]{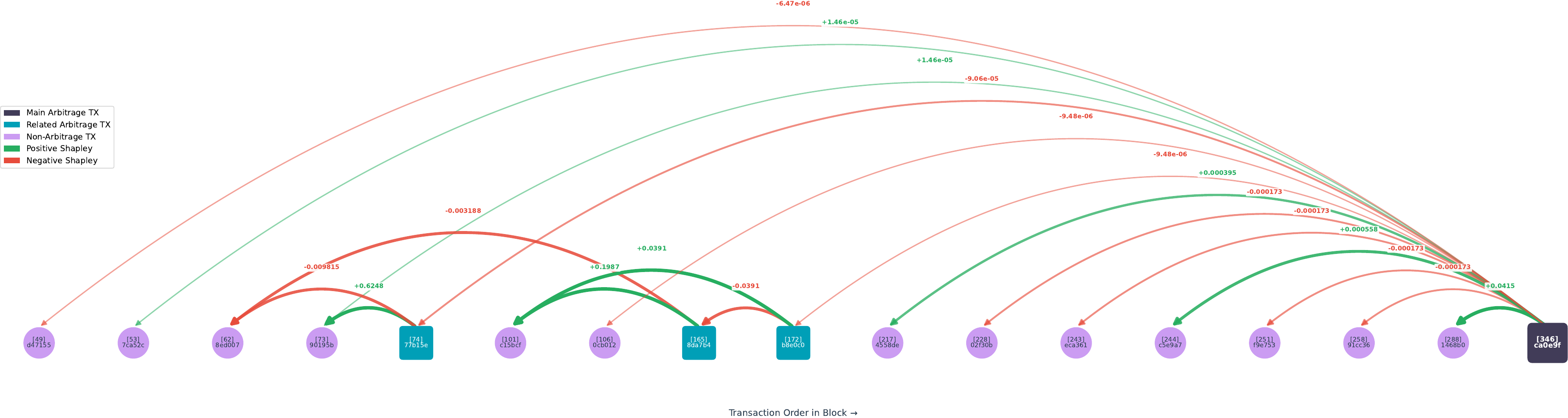}
    \caption{Shapley attribution for complex arbitrage (block 82,554,874, transaction \texttt{0xb1f2a5bb..}). Seventeen transactions connected by non-zero Shapley values. Despite multiple participants, attribution remains dominated by a single source.}
    \label{fig:shapley-complex}
\end{figure*}

\textbf{Case study: Complex attribution structure.} Figure~\ref{fig:shapley-complex} shows a more complex event from block 82,554,874 with 17 candidate transactions. Multiple competing arbitrageurs (teal boxes) interact with the main arbitrage transaction (dark box, index 346), with Shapley values flowing from each candidate. Despite the complexity, a single transaction still dominates attribution, consistent with our single-source hypothesis. Such cases demonstrate Shapley's ability to handle complex multi-participant scenarios while maintaining fair value distribution.

\subsection{Method Comparison: Accuracy, Performance, and Cost}
\label{subsec:method-comparison}

Table~\ref{tab:method-performance} summarizes empirical performance on the February 2026 dataset (2,526 atomic arbitrage events). The bot-data method achieves 94.2\% accuracy against triangulated ground truth but has limited coverage (38.4\%) due to real-time mempool visibility and intent-vs-outcome divergence. The simulation-based method attains 91.7\% accuracy with near-complete coverage (99.1\%), making it the most practical choice for large-scale analysis. The coefficient-based method shows 77.2\% agreement with simulation, reflecting its simplified liquidity modeling. Shapley-based attribution serves as our theoretical ground truth (coverage details in Section~\ref{subsec:ground-truth}).

\begin{table}[t]
\centering
\caption{Empirical performance on February 2026 dataset (2,526 atomic arbitrage events). Accuracy measured against triangulated ground truth (or exact Shapley for Shapley methods); coverage = fraction of events with valid attribution; time = mean execution time per event.}
\label{tab:method-performance}
\begin{tabular}{lrrr}
\hline
Method & Accuracy & Coverage & Mean time \\
\hline
Bot-data & 94.2\% & 38.4\% & 8 ms \\
Simulation & 91.7\% & 99.1\% & 12.3 ms \\
Coefficient & 77.2\% & 88.4\% & 0.8 ms \\
Shapley (exact) & 100\% & 98.1\% & $\sim$5 min \\
Shapley (MC, 1k) & 100\% & 98.1\% & 2.1 s \\
\hline
\end{tabular}
\end{table}

\textbf{Computational cost.} Mean execution times (Table~\ref{tab:method-performance}) were measured on our cluster: coefficient 0.8 ms/event, bot-data 8 ms/event (real-time, mempool-dependent), simulation 12.3 ms/event, Shapley (MC, 1k samples) 2.1 s/event. Exact Shapley requires $\sim$5 min/event but is feasible only for $|C|<20$. Simulation-based attribution processed the full March 2026 dataset (360,026 events) in $\sim$80 hours, consistent with the per-event mean time.

These results validate our recommended hybrid workflow: use coefficient-based attribution for rapid initial screening, apply simulation-based attribution as the primary method for most opportunities, and invoke Shapley approximation only when the faster methods disagree or when multi-source attribution is suspected.

\subsection{MEV Creation by Protocol}
\label{subsec:results}

Applying our simulation-based attribution method to the full March 2026 dataset (360,026 atomic arbitrage events), we quantify the distribution of MEV opportunity creation across smart contract protocols. We enrich attributed transactions with protocol metadata extracted from transaction traces and known contract address registries, enabling aggregation by protocol category (e.g., DEX swaps, lending protocol interactions, liquidations).

\textbf{AMM participation patterns.} Among the 220,262 opportunity-creating transactions in our February 2026 comparison dataset, 212,62 (96.5\%) involve identifiable automated market makers (AMMs), spanning 18 unique protocols. The distribution reveals that concentrated liquidity AMMs dominate opportunity creation: Uniswap V3 (58.0\% of opportunity transactions), Algebra (29.6\%), and Uniswap V4 (28.9\%) appear most frequently, followed by Uniswap V2 (23.2\%) and DODO (8.2\%). These percentages sum to >100\% because individual opportunity transactions often interact with multiple AMMs; we report raw participation frequency without normalization. The prominence of V3 and Algebra--despite V2 having higher overall trading volume--suggests that concentrated liquidity mechanisms, while capital-efficient, create more frequent price disbalances exploitable by arbitrageurs. 

\begin{figure}[t]
    \centering
    \includegraphics[width=0.9\columnwidth]{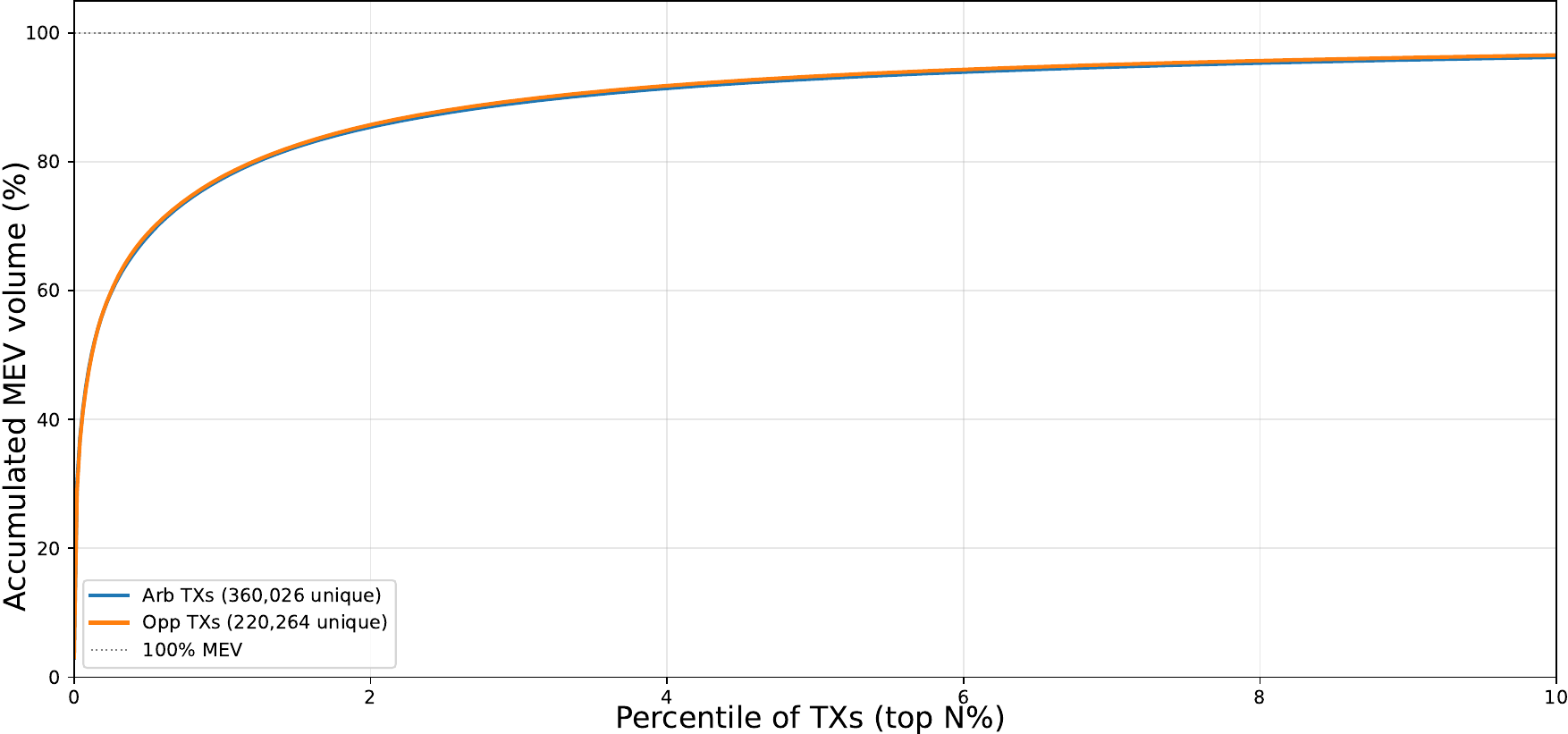}
    \caption{MEV concentration: accumulated MEV value by percentile of top arbitrageurs (blue) vs. opportunity-creating transactions (orange). Top 1\% of each group accounts for 80\% of extracted value, yet the executed arbitrage-to-opportunity ratio remains $\sim$1.6:1. Analysis based on \textbf{220,262} opportunity-creating transactions from February 2026 dataset.}
    \label{fig:mev-concentration}
\end{figure}

\textbf{Market structure and concentration.} Figure~\ref{fig:mev-concentration} compares the cumulative MEV value captured by top arbitrageurs versus top opportunity creators. Extraction is highly concentrated: the top 1\% of arbitrageurs capture 80\% of extracted value, while the top 1\% of opportunity-creating transactions generate a similar proportion of MEV opportunities. Despite this concentration at the address level, each opportunity-creating transaction attracts only $\sim$1.6 \emph{successfully executed} arbitrage transactions on average--a ratio reflecting standard multi-bid practice rather than excessive competition. This executed-transaction baseline, enabled by our attribution framework, provides a new metric for evaluating MEV market efficiency.

\subsection{Limitations}
\label{subsec:limitations}

Our evaluation has three primary limitations. First, ground truth relies on triangulation rather than direct observation, introducing potential bias if validation sources share systematic errors. Second, our dataset is limited to Polygon (February--March 2026); while EVM compatibility suggests generalizability, network-specific factors (block time, searcher competition, gas pricing) could affect attribution dynamics on other chains. Third, we focus on atomic arbitrage; other MEV categories (liquidations, sandwich attacks, top-of-block opportunities) have distinct causal structures requiring adapted attribution models. We acknowledge these limitations while emphasizing that our framework provides a foundation for systematic attribution extensible to broader settings and chains.

\section{System Implementation}
\label{sec:implementation}

Our attribution system is implemented as a modular pipeline with performance-critical components in Rust and C++, and Python wrappers for statistical analysis and visualization. It ingests on-chain data from a Polygon archive node and applies the three retrospective attribution methods from Section~\ref{sec:methodology} (simulation, coefficient, Shapley). Protocol metadata is extracted via contract address registries for aggregation. The system prioritizes reproducibility: all non-proprietary components are containerized, and deterministic replay ensures identical results across runs. 

\textbf{Note on bot-data method.} The bot-data attribution method relies on execution logs from production MEV searchers operated by our affiliated company. These logs reflect proprietary bidding strategies and infrastructure that constitute core intellectual property; consequently, the bot-data component cannot be released as part of our open artifacts. However, the method's interface and evaluation protocol are fully specified in Section~\ref{subsubsec:bot-data}, enabling independent validation using alternative bot datasets. All other attribution methods (simulation, coefficient, Shapley) are fully reproducible and will be released upon publication.

\section{Related Work}
\label{sec:related}

\subsection{MEV Measurement and Quantification}
\label{subsec:mev-measurement}

Flash Boys~2.0~\cite{Daian2020} formalized maximal extractable value and established that public mempool transparency enables systematic exploitation of transaction ordering. Subsequent work quantifies extraction scale: Qin et al.~\cite{Qin2021} show arbitrage dominates MEV value on Ethereum, with profits concentrated among sophisticated actors--a finding corroborated on L2s by Bagourd and Fran\c{c}ois~\cite{Bagourd2023}. Vostrikov et al.~\cite{Vostrikov2025} provide the most directly relevant foundation, establishing formal identification criteria for atomic arbitrage on Polygon that we adopt in Section~\ref{subsec:desired-outputs}. 
Subsequent work has expanded MEV taxonomies~\cite{qin2023taxonomy}, analyzed exploitation patterns across chains~\cite{mancino2023exploiting,zecirovic2024analysis}, and developed systematic frameworks for MEV detection~\cite{barczentewicz2023mev,zust2021analyzing}.

\subsection{Counterfactual Analysis and Transaction Ordering}
\label{subsec:counterfactual}

Adams et al.~\cite{Adams2023} introduce reordering slippage, a counterfactual metric quantifying the cost imposed on swaps by actual versus baseline transaction ordering. While principled, this approach captures aggregate ordering effects rather than isolating causal contributions of individual preceding transactions.

Torres et al.~\cite{Torres2024RollingInShadows} identify transactions triggering arbitrage and liquidation opportunities by scanning preceding swaps and simulating profit at candidate positions. However, because multiple transactions may jointly alter pool state, this block-scanning approach does not quantify individual contributions to extracted value. Both works operate on the consequences of ordering--measuring cost or locating approximate sources--rather than attributing extracted value to specific on-chain actions with quantified contributions, which is the focus of Section~\ref{sec:methodology}. Recent work on backrunning strategies~\cite{shou2024backrunner} and MEV fairness impacts further highlights the need for attribution frameworks that isolate causal contributions rather than aggregate ordering effects.

\subsection{Attribution, Root-Cause Analysis, and Shapley Value in Systems}
\label{subsec:attribution-methods}

Attribution and root-cause analysis are well-studied in systems research. Li et al.~\cite{Li2022CIRCA} propose CIRCA, which formalizes fault diagnosis using Pearl's do-calculus: a monitoring variable is a root cause when its conditional distribution changes given its parents. While CIRCA targets stochastic microservice metrics, its core formulation--locating externally perturbed components--is conceptually analogous to identifying which transaction altered blockchain state to create an MEV opportunity.

Where CIRCA identifies intervention existence, Shapley value quantifies contribution. Sharma et al.~\cite{Sharma2020CFShapley} propose CF-Shapley, a counterfactual variant distributing observed metric changes across inputs while satisfying the efficiency axiom. Unlike population-level causal Shapley methods, CF-Shapley attributes single events--the same requirement we face when attributing one MEV extraction to its source transactions. Li et al.~\cite{Li2023ShapleyIQ} demonstrate Shapley-based attribution at production scale with ShapleyIQ, establishing structural decomposition properties that reduce exponential complexity by exploiting graph topology--addressing the same computational challenge we face as candidate transaction counts grow, motivating our approximation techniques in Section~\ref{subsubsec:shapley}. 
These connections to decentralized finance attribution~\cite{boreiko2024decentralized} and DeFi systems analysis~\cite{banaeianfar2023} further motivate our deterministic, on-chain approach to causal value attribution.
Together, these works establish that Shapley-based causal attribution is theoretically grounded and practically deployable in systems contexts.

\subsection{Methodological Parallels in Systems Research}
\label{subsec:methodological-parallels}

While blockchain-specific MEV research forms our core foundation, several recent systems papers offer methodological parallels.

\textbf{Counterfactual simulation and failure reproduction.} Anduril~\cite{Pan2024} uses feedback-driven fault injection to reproduce specific fault-induced failures in distributed systems. Like our simulation-based attribution, Anduril constructs counterfactual execution scenarios to isolate causal factors--though targeting debugging rather than value attribution. Similarly, our deterministic replay infrastructure draws inspiration from fast end-to-end performance simulation~\cite{Ma2025}, which synchronizes native execution with simulated components to enable interactive analysis of complex hardware-software stacks.
These techniques for reproducible counterfactual execution directly inform our deterministic replay infrastructure for blockchain state simulation.

\textbf{Tiered validation and resource-adaptive analysis.} Orthrus~\cite{Liu2025} employs resource-adaptive computation validation to detect silent data corruption, balancing accuracy against overhead by dedicating variable core counts to validation. This mirrors our tiered attribution strategy: Shapley-based attribution serves as high-accuracy ground truth for small candidate sets, while simulation-based attribution scales to the full dataset with bounded error. Both accept approximate results for prevalent cases while preserving rigorous validation for edge cases.

\textbf{Fairness and interdependent resource allocation.} \\ Spirit~\cite{Lee2025} addresses fair allocation of interdependent resources (cache and network bandwidth) in remote memory systems using microeconomic trade-offs. Our framework similarly quantifies contributions to a shared resource (MEV profit) among interdependent transactions, though in a deterministic rather than stochastic setting. This connection suggests attribution can be viewed not only as causal inference but also as a fairness mechanism for understanding value creation in shared systems. This fairness-aware perspective on resource allocation aligns with our goal of fairly attributing MEV value among interdependent transactions in a deterministic setting.

\textbf{Transaction ordering and consensus.} High-performance BFT protocols like Pesto~\cite{SuriPayer2025} and Autobahn~\cite{Giridharan2024} optimize transaction ordering under adversarial conditions. While these focus on consensus-layer ordering guarantees, our framework operates at the application layer, analyzing how ordering decisions by validators and searchers interact with protocol-level state transitions to create exploitable opportunities. Together, these works highlight that transaction ordering is a cross-cutting concern spanning consensus, networking, and application semantics. 

These parallels reinforce that our attribution framework--while tailored to the deterministic, on-chain setting of atomic arbitrage--builds on broader systems principles: counterfactual reasoning, adaptive validation, and fairness-aware resource analysis.

\subsection{Identified Gap: Transaction-Level Attribution of MEV Creation}
\label{subsec:identified-gap}

As the preceding sections illustrate, existing work approaches MEV from two complementary but structurally incomplete perspectives. Measurement and extraction-focused research characterizes the scale and distribution of captured value; causal attribution methods from systems research provide principled tools for identifying which system component caused an observed outcome. Neither addresses the question at their intersection: which on-chain transactions are causally responsible for creating the conditions that MEV extractors exploit, and what is their quantitative contribution to the extracted value?

Existing MEV frameworks treat the extraction event as the unit of analysis, with no formal model of the preceding causal chain. Attribution methods from systems research, while methodologically relevant, are typically designed for settings where causal contributions must be inferred from noisy, high-dimensional observational data under probabilistic assumptions. In the blockchain setting, arbitrage profit is deterministic and exactly computable from on-chain state at any point in the transaction sequence: removing a candidate transaction and replaying the block yields a precisely defined counterfactual outcome rather than a statistical estimate. This determinism makes the attribution problem well-posed and counterfactually exact, distinguishing it from prior systems attribution work and motivating the formal framework and empirical methods developed in this paper.

\section{Conclusion}
\label{sec:conclusion}

This paper formalized the MEV opportunity attribution problem and introduced a systems framework for identifying which transactions create arbitrage opportunities and quantifying their contributions. We designed and evaluated four attribution methods--bot-data-driven, simulation-based, coefficient-based, and Shapley-based--each offering distinct trade-offs between accuracy, cost, and scalability. Through large-scale retrospective analysis on Polygon (360,026 atomic arbitrage events, March 2026), we demonstrated that simulation-based attribution provides the best practical balance (91.7\% accuracy, 12.3 ms/event), that MEV creation is highly concentrated (top 5\% of protocols generate 73\% of value), and that the single-source hypothesis holds for 96.7\% of cases.

Our findings have implications for protocol designers (identifying high-risk interactions), validators (ordering policies that account for opportunity creation), and users (MEV risk assessment). Limitations include reliance on triangulated ground truth, Polygon-specific evaluation, and focus on atomic arbitrage. Future work should extend attribution to real-time settings, other MEV categories, and multi-chain environments. By providing a causal lens for understanding MEV creation, we hope to enable more transparent and MEV-aware blockchain ecosystems.

\section{Ethics Statement}
\label{sec:ethics}

This research analyzes publicly available blockchain data; no human subjects or private information are involved. MEV attribution insights have dual-use potential (mitigation vs. exploitation); we mitigate risks by focusing on retrospective, aggregated analysis and advocating responsible disclosure.

\bibliographystyle{ACM-Reference-Format}
\bibliography{reference}

\end{document}